\documentclass[conference]{IEEEtran} 
\usepackage{amsmath,amssymb}
\usepackage{bm}
\usepackage{mathtools}
\usepackage{steinmetz}
\usepackage{graphicx}
\usepackage{epstopdf}
\usepackage{setspace} 
\usepackage{multirow}
\usepackage{mwe}
\usepackage{algpseudocode}
\usepackage[ruled]{algorithm2e}

\usepackage{verbatim} 
\usepackage{subfigure}

\SetKwInput{KwInput}{Input}               
\SetKwInput{KwOutput}{Output}
\begin{document}
\pagestyle{empty}    
\title{Intelligent Reflecting Surface Aided Vehicular Communications}
\author{
\IEEEauthorblockN{Dilin~Dampahalage,~K. B. Shashika Manosha,~Nandana~Rajatheva,~
        and~Matti~Latva-aho} \\
        \IEEEauthorblockA{\textit{Centre for Wireless Communications,} 
\textit{Univeristy of Oulu, Finland}\\
E-mail: \{dilin.dampahalage, nandana.rajatheva, matti.latva-aho\}@oulu.fi, manoshadt@gmail.com}
}

\maketitle
\thispagestyle{empty}
\begin{abstract}
We investigate the use of an intelligent reflecting surface (IRS) in a millimeter-wave (mmWave) vehicular communication network. An intelligent reflecting surface consists of passive elements, which can reflect the incoming signals with adjustable phase shifts. By properly tuning the phase shifts we can improve link performance. This is known as phase optimization or passive beamforming. We consider the problem of rate maximization in the uplink, which utilizes an IRS. However, using an IRS brings more challenges in terms of channel estimation. We propose two schemes to reduce the channel estimation overhead associated with utilizing an IRS. One method uses the grouping of reflecting elements and the other one performs passive beamforming based on the position of the device. Numerical results show IRS can bring significant improvements to existing communication. Furthermore, to get a practical insight into vehicular communications aided by an IRS, we use a commercial ray-tracing tool to evaluate the performance. 
\end{abstract}


\section{Introduction} \label{introduction}
The wireless propagation environment is random and uncontrollable. Recently intelligent reflecting surfaces have been proposed as a means of having some control over it with the help of software-controlled reflections \cite{wu2019towards}. An IRS consists of a planar array of passive reflecting elements that can reflect the incoming rays with adjustable phase shifts and gains. A reflecting element in the IRS captures the incoming signal and re-scatters it in every direction \cite{basar2020indoor}. The effective path loss is the product of the path losses of individual links through the IRS. The reflecting elements in the IRS can work together to achieve fine-grained three-dimensional reflect beam forming \cite{wu2019towards}. A survey on IRSs is presented in \cite{zhao2019survey}.


The effective channel through the IRS consists of the direct channel from the base station (BS) to the user, indirect channel from BS to IRS, and reflection channel from IRS to the user. Most theoretical work on IRSs assumes that perfect channel state information (CSI) is available at the IRS for the indirect and reflection channels. However, acquiring the CSI for the indirect and reflection channels is challenging since the reflecting surface consist of passive elements and limited processing power. Therefore, it is important to consider robust transmission strategies for IRS aided wireless systems. Robust active precoder and passive reflection beamforming design is investigated in \cite{robust_beamforming_irs} based on the assumption of an imperfect reflection channel. Robust transmission design based on imperfect cascaded BS-IRS-user channels is presented in \cite{robust_transmission_irs}.

We can utilize IRSs to improve communication over existing communication networks. The transmit beamforming must be jointly optimized with the reflect beamforming at the IRS based on BS-IRS, IRS-user, and BS-user channels, to fully reap the network beamforming gain. An efficient algorithm to tackle this optimization problem based on alternating optimization of the phase shifts and transmit beamforming vector in an iterative manner is proposed in \cite{wu2019intelligent}. This is further extended in \cite{8930608} to consider discrete phase shifts at the IRS, which is how the practical implementation can be done. Here the problem is to minimize the transmit power required at the access point (AP) by jointly optimizing the beamforming at the AP and discrete phase shifts at the IRS. Unfortunately, this is an NP-hard problem. However, they propose a sub-optimal algorithm with zero-forcing (ZF) based linear precoding at the AP for low-complexity implementation.

One of the most recent applications of IRSs is symbiotic radio transmission, where the relevant information bits are transferred by the on/off states of the IRS \cite{yan2019passive}, \cite{zhang2020large}. An unmanned aerial vehicle (UAV)-assisted IRS symbiotic radio system is studied in \cite{hua2020uavassisted} where multiple IRSs are available to sense the environmental information in an urban environment. Recent studies also focus on utilizing multiple IRSs. Weighted sum rate (WSR) maximization of the cell edge users utilizing multiple IRSs is considered in \cite{wsr_multi_irs}. WSR maximization of an IRS assisted simultaneous wireless information and power transfer (SWIPT) multiple-input multiple-output (MIMO) system is studied in \cite{w_pow_trans}. The enhancement of the performance of a multi-cell MIMO communication system utilizing IRSs is investigated in \cite{multicell_irs}. 

%
Vehicular communication networks have been studied extensively to realize the concept of intelligent transportation systems (ITS) \cite{8594703}. Various vehicular applications are been considered, including a wide variety of safety-oriented, comfort, and entertainment applications \cite{8539687}. These applications bring great challenges to existing communication and networking technologies. The introduction of data-intensive sensors such as laser imaging detection and ranging (LIDAR) has resulted in vehicular communication networks having to support Gb/s data rates \cite{8539687}. Large system bandwidth is needed for such high data rates. These facts have motivated to utilize the mmWave frequency band (10 GHz-300 GHz) that has a largely available bandwidth \cite{8539687} for vehicular communications. However, mmWave frequencies experience a higher path loss, thus reducing the transmission range. Recently, IRSs have been proposed in \cite{wang2019intelligent} to overcome these challenges. They consider a scenario where multiple IRSs are deployed to assist mmWave communications. The received power is maximized by jointly optimizing transmit beamforming at the BS and passive beamforming at the IRS.

The use of IRSs to assist a vehicular network has been briefly considered in the literature. An IRS enabled vehicular network is considered in \cite{makarfi2020reconfigurable} to improve the physical layer security. Two vehicular network models have been presented in \cite{makarfi2020reconfigurable}. One model with an IRS based AP and another model with an IRS based relay. Analysis of outage in IRS assisted vehicular networks is considered in \cite{wang2020outage}. The outage performance of traditional vehicular networks and IRS aided vehicular networks are compared \cite{wang2020outage}. However, IRSs bring new challenges in terms of channel estimation, even more with high mobility in vehicular networks. Estimating the channel coefficients for each reflecting path takes a large overhead. Although the rate maximization problem for IRS aided systems has been considered in various studies in the literature \cite{perovi2019channel}, \cite{xiu2020irsassisted}, to the best of the authors' knowledge this is the first work that considers this problem for vehicular communication.

In this paper, we consider a mmWave vehicular communication system facilitated by an IRS. We consider the problem of rate maximization in the uplink, where an IRS is used to improve the link performance. We use a successive refinement \cite{8930608} based algorithm to optimize the IRS phase shifts. We propose two schemes for IRS phase optimization that reduce the channel estimation overhead and facilitate the use of large reflecting arrays. One method is based on the grouping of reflecting elements in the IRS. The other method utilizes position-based passive beamforming. Numerical results show that the achievable rate can be significantly improved by using an IRS. Further,  we evaluate the performance of the system under mobility with a commercial ray-tracing tool, Wireless InSite \cite{insite}. The results further validate that large reflecting arrays can provide significant benefits over existing communication.

The rest of the paper can be summarised as follows. The system model and problem formulation are presented in Section \ref{system}. The phase optimization algorithm and two phase optimization schemes are proposed in Section \ref{schemes}. Finally, the results and conclusion are presented in Sections \ref{results} and \ref{conclusion}.

\section{System Model and Problem Formulation} \label{system}
\begin{figure}[ht]
\centerline{\includegraphics[width=8cm]{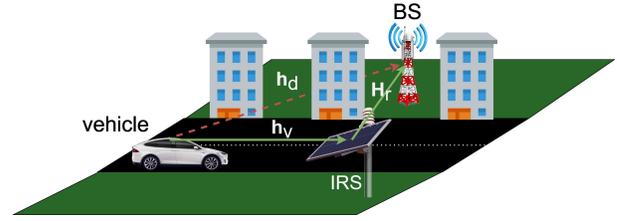}}
\caption{An illustration of IRS-aided vehicular communication network.}
\label{fig:architecture}
\end{figure}

This section presents the system model and problem formulation. Here, we consider a vehicular network consisting of a BS with $M$ antennas and a vehicle with a single antenna, and focus on the uplink. An IRS with $N$ reflecting elements is used to assist the communication. An illustration of this system is shown in Fig. \ref{fig:architecture}. Let $\bm{H}_r \in \mathbb{C}^{M \times N}$ be the channel between IRS and BS, $\bm{h}_v \in \mathbb{C}^N$ be the channel between the  vehicle and IRS, and $\bm{h}_d \in \mathbb{C}^M$ be the direct channel between the vehicle and BS.

Reflecting elements of the IRS can induce phase shifts in the incoming signals. We consider these phase shifts to be one of $L$ discrete levels. For simplicity, we assume these phase shifts take one of the values obtained by uniformly quantizing the interval $[0,2\pi)$. Thus, the set of discrete phase shifts at each reflecting element is given by
\begin{equation}
    \mathcal{F}=\left\{0,\Delta\theta,...,(L-1)\Delta\theta\right\},
\end{equation}
where $\Delta\theta=2\pi/L$. Let the, phase shifts of $i$th reflector of IRS be, $\theta_i \in \mathcal{F}$. We denote the reflection matrix of the IRS by $\bm{\Theta}$, where  $\bm{\Theta}=\text{diag}([\exp({j\theta_1}),\exp({j\theta_2}),...,\exp({j\theta_N})]^T)$.

The received signal at the BS is
\begin{equation}
    \bm{y} = (\bm{h}_d + \bm{H}_r \bm{\Theta} \bm{h}_v)x + \bm{n},
    \label{eq:1}
\end{equation}
where $x$ denote the transmit signal of the vehicle and $\bm{n}=[n_1,n_2,...,n_M]$ with $n_m \sim \mathcal{CN}(0, N_0)$ being the complex additive white Gaussian noise (AWGN) at BS antennas. The BS applies a linear beamforming vector $\bm{w} \in \mathbb{C}^M$ to decode $x$, i.e.,
\begin{equation}
    \hat{y} = \bm{w}^H(\bm{h}_d + \bm{H}_r \bm{\Theta} \bm{h}_v)x + \bm{w}^H\bm{n}.
    \label{eq:2}
\end{equation}
For simplicity we assume maximal-ratio combining (MRC) at the BS with $\bm{w}=\frac{\bm{h}_d + \bm{H}_r \bm{\Theta} \bm{h}_v}{\|\bm{h}_d + \bm{H}_r \bm{\Theta} \bm{h}_v\|}$. The signal to noise ratio (SNR) can be written as,
\begin{equation}
    \text{SNR} = \frac{P{\|\bm{h}_d + \bm{H}_r \bm{\Theta} \bm{h}_v\|}^2}{N_0},
    \label{eq:3}
\end{equation}
where $P$ is the transmit power. Using (\ref{eq:3}), the achievable rate can be expressed as,
\begin{equation}
    \mathnormal{R} = \log_2\left(1 + \frac{P{\|\bm{h}_d + \bm{H}_r \bm{\Theta} \bm{h}_v\|}^2}{N_0}\right) \text{bit} / \text{s} / \text{Hz}.
    \label{eq:4}
\end{equation}

As we can see from (\ref{eq:4}), the achievable rate is dependent on the reflection matrix at the IRS. Therefore it should be possible to achieve higher rates by properly tuning the reflection matrix. This is known as passive beamforming or phase optimization. We consider the phase optimization problem at the IRS, where the achievable rate is maximized by choosing the optimum discrete phase shifts:
\begin{equation}
\begin{aligned}
\text{maximize} \quad & \mathnormal{R} \\
\text{subject to} \quad & \theta_i \in \mathcal{F},\text{ for i=1,2,...,N},
\end{aligned}
 \label{eq:5}
\end{equation}
where $\theta_i$ $\text{for all i=1,2,...,N}$ are the optimization variables.

\section{Algorithm Development} \label{schemes}
 The optimization problem in (\ref{eq:5}) is non-convex due to discrete phase shifts. If brute force is used, the algorithm has to go through $N^L$ possibilities, which is not feasible for large reflecting arrays. However, it is possible to come up with a simple algorithm based on successive refinement \cite{8930608} by expanding the channel gain expression. Let us define $\bm{\Phi}=\bm{H}_r \text{diag}(\bm{h}_v)$ and $\bm{v}=[\exp({j\theta_1}),\exp({j\theta_2}),...,\exp({j\theta_N})]^T$. Also, let $\bm{A}=\bm{\Phi}^H\bm{\Phi}$ and $\bm{b}=\bm{\Phi}^H\bm{h}_d$. The channel gain is given by,
\begin{equation}
{\| \bm{h}_d + \bm{H}_r \bm{\Theta} \bm{h}_v\|}^2 = \bm{v}^H \bm{A} \bm{v} + 2 \mathit{Re} \{ \bm{v}^H \bm{b} \} + {\| \bm{h}_d \|}^2.
\label{eq:7}
\end{equation}

We can focus on a single reflecting element, $v_n$ considering all the other reflecting elements ($v_i$, $i\neq n$) fixed. We can write the channel gain as,
\begin{equation}
    2\mathit{Re}\{v_n^*\kappa_n\} + \tau_n,
    \label{eq:8}
\end{equation}
where $\kappa_n=\sum_{j \neq n}A_{nj}v_j + b_n$ and $\tau_n=\sum_{j \neq n}\sum_{i \neq n}v_i^*A_{ij}v_j +  2\mathit{Re}\{\sum_{i \neq n}v_i^*b_i\} + A_{nn} + \|\bm{h}_d\|^2$. Here, $A_{ij}$ and $b_i$ represent the individual elements of $\bm{A}$ and $\bm{b}$ respectively.

Based on the channel gain expression in (\ref{eq:8}) we can maximize the channel gain by matching the reflecting array phase shift $v_n$ to the phase of $\kappa_n$. For discrete phase shifts, we can formulate the successive refinement algorithm as presented in Algorithm \ref{alg:1} \cite{8930608}. Here, $\epsilon$ is the stop threshold for the convergence.

To perform passive beamforming we need to estimate all the channels involved. In addition to the direct channel between the vehicle and BS, there are individual channels through each reflecting element of the IRS. An IRS generally consists of a large number of reflecting elements. This makes the channel estimation very challenging. In the next two subsections, we propose two phase optimization schemes that can be followed to reduce the channel estimation overhead. 

\begin{algorithm}
	initialize $\bm{\Theta} = \bm{\Theta}^{(0)}$\\
	$\mathnormal{R}^{(0)}=0$\\
	set $k=1$\\
	$\mathnormal{R}^{(k)} = \log_2\left(1 + \frac{P{\|\bm{h}_d + \bm{H}_r \bm{\Theta} \bm{h}_v\|}^2}{N_0}\right)$\\
	\While{$|\mathnormal{R}^{(k)}-\mathnormal{R}^{(k-1)}|>\epsilon$}{
		\For{n=1 to N}{
			$\theta_n^*= \arg \min_{\theta \in \mathcal{F}} |\theta - \phase{\kappa_n}|$\\
		}
		$k=k+1$\\
		$\mathnormal{R}^{(k)} = \log_2\left(1 + \frac{P{\|\bm{h}_d + \bm{H}_r \bm{\Theta} \bm{h}_v\|}^2}{N_0}\right)$\\
	}
	\caption{Successive Refinement algorithm}
	\label{alg:1}
\end{algorithm}

\subsection{Grouping based IRS Phase Optimization}
One approach to reducing the channel estimation overhead is to divide the reflecting array into subgroups and perform phase optimization per subgroup. Here, all the elements in the subgroup are considered as a single element. Therefore, we only need to estimate the channel for each group, not for all the reflecting elements. Also, we can run the phase optimization algorithm considering subgroups, without going through all the individual elements.

\begin{figure}[h]
\centerline{\includegraphics[width=8cm]{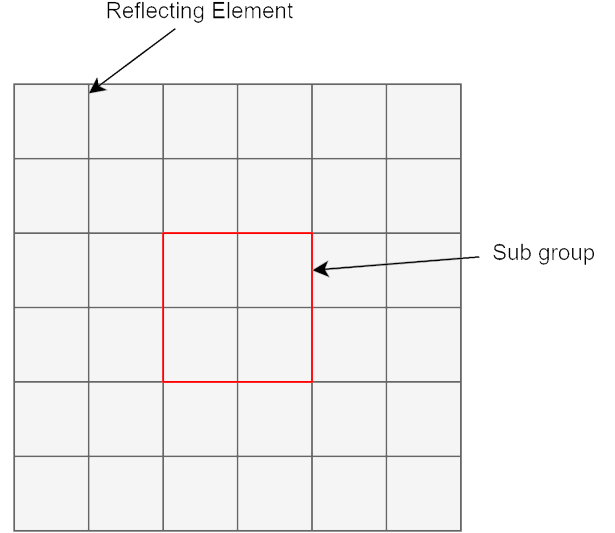}}
\caption{Reflect array divided into subgroups.}
\label{fig:subgroups}
\end{figure}

For example, Fig. \ref{fig:subgroups} shows a $6 \times 6$ reflecting array. It is divided into $2 \times 2$ sized subgroups. Each subgroup is considered as a single reflecting element. Therefore, passive beamforming is effectively done for a $3 \times 3$ reflecting array. This reduces the channel estimation overhead as well as the complexity of the successive refinement algorithm. After the phase shifts are found, phase shifts of all the reflectors in the group are set to the same value for the subgroup.
\subsection{Position based Passive Beamforming}
It is known that compared to the change of CSI in the mmWave band, the directional angles change much slower \cite{angle_based_beamforming}. Also, the beam direction of a vehicle undergoing vehicle to infrastructure (V2I) communication can be determined based on the position and the velocity of the vehicle \cite{mmv2x}.  The communication mostly happens through line-of-sight (LOS) links in this scenario \cite{mmv2x}. Therefore, we can assume that the network tracks the position of the device and also, the departure and arrival angles. Then we can determine the LOS channel matrix. Assuming planar arrays at both the BS and IRS, the LOS channel between BS and IRS can be expressed as,
 \begin{equation}
    \bm{H}_{r,los} = \sqrt{L_{los}}\exp\left(\frac{-j 2\pi d}{\lambda}\right)\bm{a}_{bs}(\phi^{bs},\theta^{bs})\bm{a}^H_{irs}(\phi^{irs},\theta^{irs}),
    \label{eq:9}
\end{equation}
where, $d$ is the distance between BS and IRS, $\bm{a}_{bs}(\phi^{bs},\theta^{bs}) \in \mathbb{C}^M$ is the array response of BS for the considered azimuth and elevation arrival angles, and $\bm{a}_{irs}(\phi^{irs},\theta^{irs}) \in \mathbb{C}^N$ is the array response of IRS for the considered azimuth and elevation departure angles. Similarly, $\bm{h}_{v,los}$ and $\bm{h}_{d,los}$ can be defined. The system can then perform passive beamforming based on the estimated LOS channels. Here, we only need to estimate the arrival and departure angles for the whole reflecting array. This takes less overhead than estimating the channel for individual reflecting elements, which is the case for full CSI. 
    
\section{Numerical Results} \label{results}
In this section we present the results of numerical simulations to validate the algorithms and to gain an insight into IRS aided vehicular communications. We initially present the results of stochastic simulations under static conditions. Later, we present the results of ray tracing based simulation results using Wireless InSite \cite{insite} that take mobility into account as well. In our simulations, we consider mmWave communications in the 28 GHz band with carrier frequency $f_c=24.2$ GHz. 

\subsection{Stochastic Simulations}

\begin{figure}[h]
\centerline{\includegraphics[width=9cm]{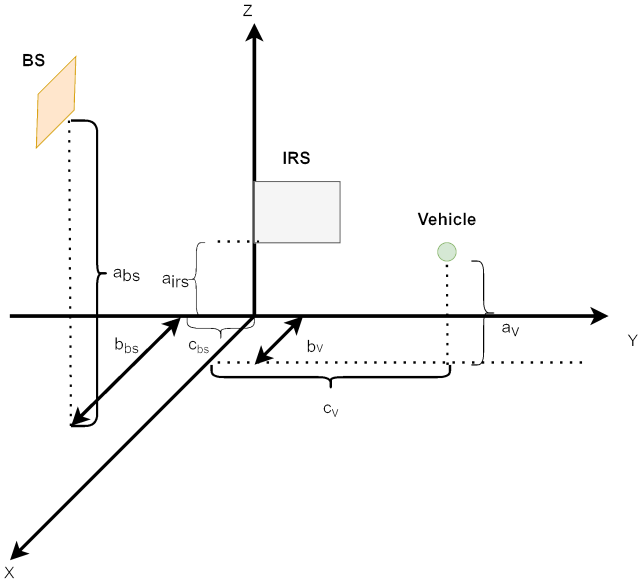}}
\caption{Device positions in the stochastic simulation.}
\label{fig:device_positions}
\end{figure}

The position of the devices in the stochastic simulation is shown in Fig. \ref{fig:device_positions}. BS has a $4\times2$ uniform planar array (UPA) antenna panel and, IRS has a planar reflecting array of $16\times16$ reflecting elements. The vehicle has a single antenna. IRS is placed on the $YZ$ plane at a height $a_{irs}=1$ m. BS is placed on the $XZ$ plane at a height $a_{bs}=2$ m and at distances of $b_{bs}=20$ m, and $c_{bs}=10$ m from the origin. Vehicle antenna is placed at a height $a_{v}=1$ m and at distances of $b_{v}=1.5$ m, and $c_{v}$ from the origin.

All the channels involved are modeled with the Rician fading model. To model path loss, 3GPP TR 38.901 UMi - Street Canyon path loss model \cite{tr38.901} is used. We use $\beta_r=2$ for IRS-BS link, $\beta_v=1$ for the vehicle-IRS link, and $\beta_d=\infty$ for the direct link, where $\beta$ denote the Rician coefficient for the associated channel.

\begin{figure}[h]
\centerline{\includegraphics[width=\columnwidth]{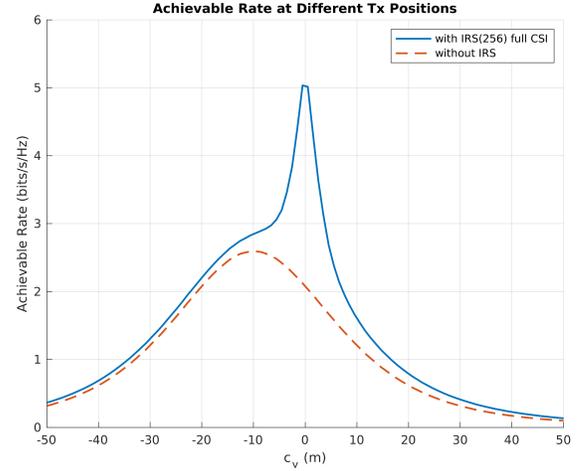}}
\caption{Variation of achievable rate as vehicle position is changed.}
\label{fig:rate_vs_pos}
\end{figure}

Fig. \ref{fig:rate_vs_pos} shows the variation of the achievable rate as we change the vehicle position by changing $c_v$. We see that the IRS gives a significant increase in the rate when the vehicle is close to the IRS. It shows the highest rate at the closest point at $c_v=0$. However, the gain in the rate decreases as the vehicle moves further away from the IRS. 

\begin{figure}[h!]
\centering
\subfigure[]{
    \includegraphics[width=\columnwidth]{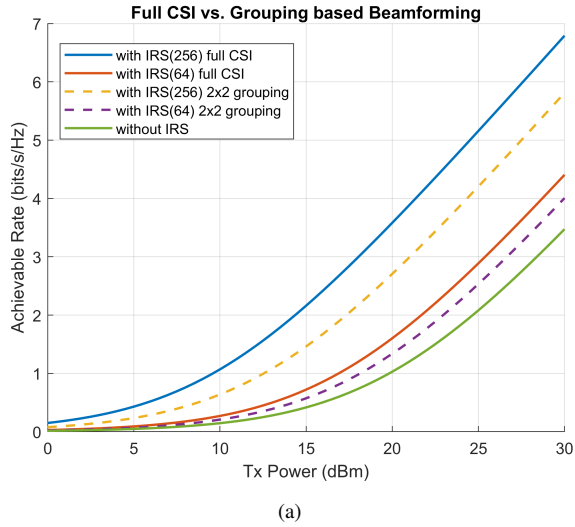}
    \label{fig:rates_sim_a}
} 
\subfigure[]{
    \includegraphics[width=\columnwidth]{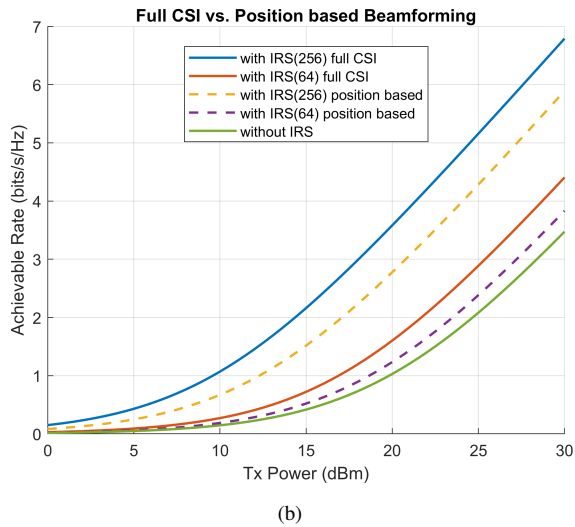}
    \label{fig:rates_sim_b}
}
\caption{Comparison of variation of achievable rate with transmit power a) with grouping and b) with position based beamforming.}
\label{fig:rates_sim}
\end{figure}

Next, we fix the vehicle position at the closest point to the IRS, which is at $c_v=0$. The achievable rate is compared for reflecting array sizes of $16\times16$ and $8\times8$ while changing the transmit power. Subgroup based phase optimization is compared in Fig. \ref{fig:rates_sim_a}. We see that there is a decrease in performance when grouping is used. However, there is a performance gain compared to the system without an IRS. When a $16\times16$ reflecting array is used with $2\times2$ grouping, it effectively acts as an $8\times8$ panel in-terms of passive beamforming. Yet, it gives better performance than utilizing an $8\times8$ reflecting array with full CSI. Therefore, grouping provides a practical means to utilize large reflecting arrays, reducing the overhead for channel estimation. 
tate   

Position based phase optimization is compared in Fig. \ref{fig:rates_sim_b}. This scheme also has a reduction in performance compared to when full CSI is available. Yet, it gives a significant increase in the rate for a $16\times16$ reflecting array compared to the system without an IRS. However, the rate of $8\times8$ reflecting array lies closely by the rate curve without IRS. This suggests that position based beamforming is more suitable for large reflecting arrays.

\begin{figure}[h!]
\centerline{\includegraphics[width=\columnwidth]{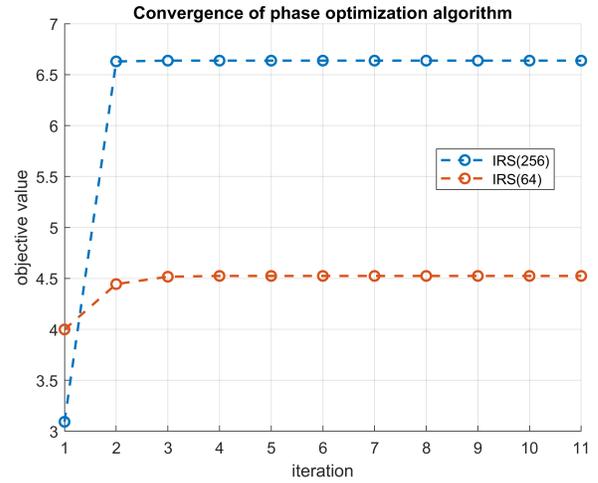}}
\caption{Convergence of the successive refinement algorithm.}
\label{fig:convergence}
\end{figure}

The convergence of the successive refinement algorithm is shown in Fig. \ref{fig:convergence}. We see that the algorithm converges quite fast for both $8 \times 8$ and $16 \times 16$  reflecting arrays. Here, we have set the initial values of phase shifts to zero ($\theta_i=0$, for all $i=1,2,..,N$).

\begin{figure}[h!]
\centerline{\includegraphics[width=\columnwidth]{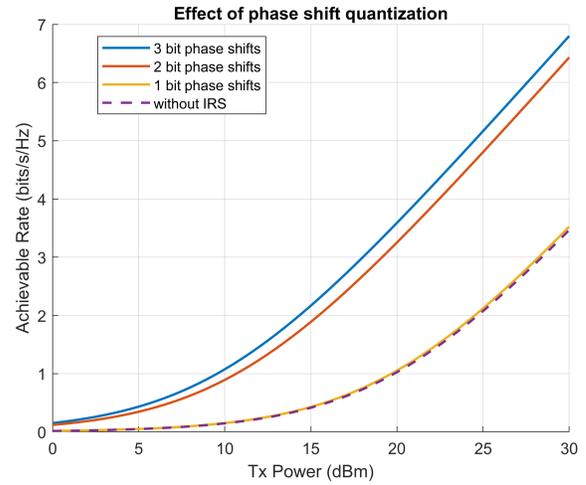}}
\caption{Illustrating the effects of quantizing the phase shifts.}
\label{fig:bit}
\end{figure}

Fig. \ref{fig:bit} illustrates the effects of quantizing the phase shifts. As we can see, 1-bit phase shifts hardly give any benefits over existing communication without IRS. As we increase discrete phase shift levels to 2 bits the performance improves dramatically. The performance gain when we further increase the number of quantization bits to 3, is not that significant. Therefore, we can conclude that a reflecting array with even 2-bit phase shifts is sufficient to improve existing communication. This is beneficial because practical implementations of IRSs can only support a limited number of phase shift levels. 

\subsection{Ray Tracing based Simulation}

\begin{figure}[h!]
\centerline{\includegraphics[width=\columnwidth]{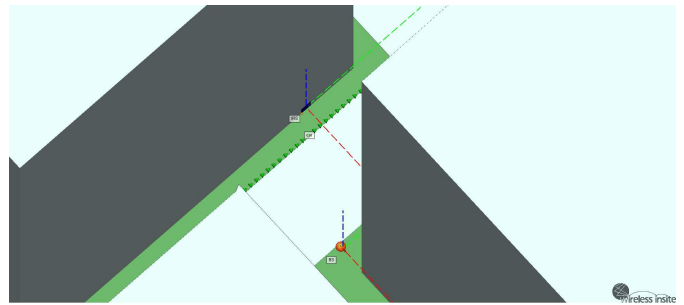}}
\caption{Simulating IRS aided system in Remcom Wireless InSite \cite{insite}.}
\label{fig:ray_tracing_scenario}
\end{figure}
Although stochastic simulations in the last section provide some insights on the IRS aided system, they rely on simplified assumptions. They do not take mobility into account either. We need a more complex simulation to model mobility and the environmental conditions in detail. Ray tracing is considered as a reliable methodology to estimate complex propagation characteristics in mmWave vehicular networks \cite{raytracing}.

\begin{figure}[h!]
\centerline{\includegraphics[width=\columnwidth]{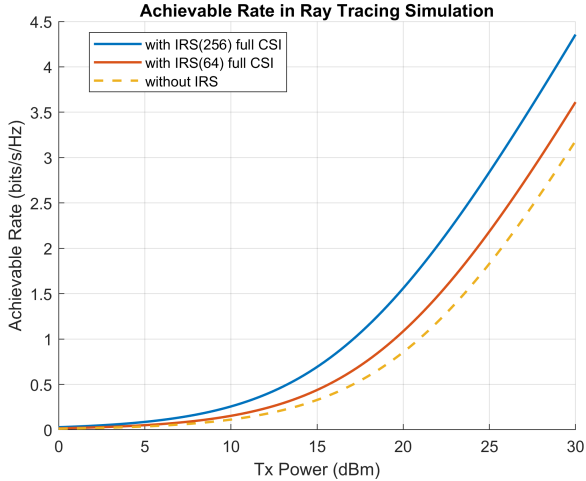}}
\caption{Achievable rate with transmit power for the ray tracing simulation.}
\label{fig:ray_tracing_rate}
\end{figure}

The IRS aided system is modeled in Remcom Wireless InSite \cite{insite} as shown in Fig. \ref{fig:ray_tracing_scenario}. The IRS is modeled as a rectangular array of patch antennas. A route for the transmitter in the vehicle is defined with an average speed of $10$ m/s. The system is modeling an urban scenario with buildings and roads. BS is positioned at one side of the road. The vehicle is moving on the other side of the road. IRS is placed closer to the vehicle's path. Vehicular traffic is not considered for simplicity. The channel matrices are obtained using the ray-tracing simulation and phase optimization is carried out. Fig. \ref{fig:ray_tracing_rate} shows the plots of achievable rates for reflecting array sizes of $8\times8$ and $16\times16$. Here, we have considered the closest point of the vehicle to the IRS and have calculated the rate while changing transmit power. A significant increase in rate is seen for the $16\times16$ reflecting array. The $8\times8$ reflecting array only provides a slight improvement over existing communication without IRS. This suggests that large reflecting arrays are needed to improve the system performance considerably. 

\section{Conclusion} \label{conclusion}
In this paper, we have analyzed the usage of an IRS in a mmWave vehicular network to maximize the achievable rate. A ray-tracing based simulation is used to validate the benefits of using an IRS in a vehicular network. We have proposed two passive beamforming schemes that can be utilized to reduce the channel estimation overhead. We have compared the performance of these methods through numerical simulations based on a stochastic channel model. The numerical results suggest, although the performance of these schemes is lower than when full CSI is available, these schemes still provide significant improvements to the existing communication. They provide practical ways to utilize large reflecting arrays.

\bibliographystyle{IEEEbib}  
\bibliography{Sec_7_Bibliography.bib}

\end{document}